\documentclass[12pt]{iopart}

\usepackage{graphics}
\usepackage{subfig}
\usepackage{color}
\begin{document}

\title[Bottle-brush co-polymers with flexible backbones under poor solvent
conditions]{Phase behaviour of two-component bottle-brush polymers
with flexible backbones under poor solvent conditions}

\author{Nikolaos G Fytas$^1$ and Panagiotis E Theodorakis$^2$}

\address{$^1$ Applied Mathematics Research Centre, Coventry University, Coventry, CV1 5FB, United Kingdom}
\address{$^2$ Department of Chemical Engineering, Imperial College London, London SW7 2AZ, United Kingdom}

\ead{nikolaos.fytas@coventry.ac.uk}
\ead{p.theodorakis@imperial.ac.uk}

\begin{abstract}

The phase behaviour of two-component bottle-brush polymers with
fully flexible backbones under poor solvent conditions is studied
via molecular-dynamics simulations, using a coarse-grained
bead-spring model and side chains of up to $N=40$ effective
monomers. We consider a symmetric model where side chains of type
A and B are grafted alternately onto a flexible backbone. The aim
of this study to explore the phase behaviour of two-component
bottle-brushes depending on parameters, such as as the grafting
density $\sigma$, the backbone length $N_b$, the side-chain length
$N$, and the temperature $T$. Based on a cluster analysis, we
identify for our range of parameters the regimes of fully phase
separated systems, i.e., A-type side chains form one cluster and
B-type chains another, while the interface that separates these
two clusters contains the backbone monomers. We find that
pearl-necklace or Janus-like structures, which normally occur for
bottle-brush polymers with rigid backbones under poor solvent
conditions, are fully attributed to the backbone rigidity, and,
therefore, such structures are unlikely in the case of bottle
brushes with fully flexible backbones. Also, a comparative
discussion with earlier work on the phase behaviour of
single-component bottle-brush polymers with flexible backbones is
performed.

\end{abstract}


\submitto{Materials Research Express}

\maketitle

\section{Introduction}
\label{introduction}

Recent advances in chemical synthesis allow for the tinkering of
macromolecules of complex architecture with well-defined
properties, suitable for exquisite
applications~\cite{Zhang2005,Sheiko2008}. Bottle-brush polymers
are macromolecules with side chains grafted onto a backbone
polymeric chain that can be intrinsically rigid or moderately
flexible due to the presence of the side chains~\cite{Zhang2005, Sheiko2008, Maleki2011,
Subbotin2007,Potemkin2009, Binder2012, Walther2013, Rathgeber2005,
Zhang2006, Theodorakis2011, Theodorakis2012, Hsu2010, Feuz2007,
Lee2008}. The interplay between steric repulsions and effective
attractions of the side-chain monomers that are grafted onto the
backbone leads to intricate spatial self-organization of these
macromolecules that can be tuned by variation of the solvent
quality, temperature, pH, etc.~\cite{Sheiko2004, Theodorakis2009,
Theodorakis2010, Theodorakis2011c, Theodorakis2011d,
Theodorakis2010b, Theodorakis2013}. Hence, bottle-brush polymers
are stimuli responsive materials that are suitable for various
applications~\cite{Zhang2005, Sheiko2008, Walther2013}.

Apart from synthetic bottle-brush polymers, also bio-polymers with
a related architecture are abundant in nature. For instance,
brush-like macromolecules that contain a protein backbone with
carbohydrate side chains known as proteoglycans~\cite{Iozzo2000},
are held responsible for a large variety of very interesting
biological functions (cell signaling, cell surface protection,
joint lubrication, etc.)~\cite{Muir1983, Kaneider2004, Klein2009}.
In view of this interest, understanding the structure-property
relation of such bottle-brush macromolecules remains a challenging
problem of statistical thermodynamics due to the multitude of the
involved length scales resulting from their complex
structure~\cite{Birshtein1984, Witten1986, Birshtein1987,
Wang1988, Ligoure1990, Ball1991, Murat1991, Dan1992, Wijmans1993,
Fredrickson1993, Li1994, Rouault1996, Sevick1996, Saariaho1997,
Saariaho1998, Rouault1998, Saariaho1999, Shiokawa1999,
Subbotin2000, Khalatur2000, Stepanyan2002, deJong2004,Elli2004,
Connolly2005, Yethiraj2006, Chang2009, Hsu2007, Hsu2008}.

Nowadays, an in-depth understanding of the structural properties
of bottle-brush polymers under $\Theta$ and good solvent
conditions has been provided by experiment, theory and computer
simulations (for instance, see the review
article~\cite{Binder2012} and references therein).
However, many applications require that these macromolecules exist
under poor solvent conditions~\cite{Walther2013}. Apart from
the experimental interest for bottle-brushes under poor
solvent conditions~\cite{Walther2013}, computer simulations
have initially considered in their studies single-component bottle
brushes under poor solvent conditions, where the backbone of the
macromolecule was strictly rigid~\cite{
Theodorakis2009, Theodorakis2010, Theodorakis2011d}. Moreover,
theory has considered both rigid and flexible bottle-brush
macromolecules~\cite{Sheiko2004}. For rigid backbones and when the
grafting density that the side chains are grafted onto the
backbone is low, the side chains do not interact with each other
and collapse individually onto the rigid backbone adopting a
globular conformation.
As the grafting density increases to intermediate densities, side
chains start to interact and clusters containing more than one
side chain form pearl-necklace structures along the backbone, in
line with theoretical predictions~\cite{Sheiko2004}. Further
increase of the grafting density results in the formation of
homogeneous brush cylinders \cite{Sheiko2004, Theodorakis2009,
Theodorakis2010, Theodorakis2011d}. When two chemically different
types of side chains are grafted alternately onto a rigid
backbone, then micro-phase separation between the different
monomers takes place leading to phase separation within pearls at
intermediate densities~\cite{Theodorakis2011c, Theodorakis2011d,
Theodorakis2010b}. At even higher densities, Janus-like structures
are seen, where one half of the cylinder contains chains of one
type and the other half side chains of the other type. The main
hypothesis of this previous work has been that the backbone is
strictly rigid disregarding in this way effects due to the
flexibility of the backbone.

Bottle-brush polymers with fully flexible backbones under poor
solvent conditions have been only recently tackled for
single-component bottle brushes~\cite{Theodorakis2013}. In this
case pearl-necklace or Janus-like structures do not occur.
Therefore, such structures can only be expected for bottle brushes
with non-flexible backbones. Here, we extend this work by studying
the phase behaviour of two-component bottle-brush macromolecules,
where two chemically different types of side chains are grafted
alternately onto a flexible backbone. We focus on understanding
the phase behaviour of A and B grafted chains and we make a
comparison with two-component bottle brushes with rigid
backbones~\cite{Theodorakis2011c, Theodorakis2011d,
Theodorakis2010b}, as well as with single-component bottle-brushes
with flexible backbones~\cite{Theodorakis2013}.

The remainder of this manuscript is organized as follows: In
section~\ref{model} we describe our model and sketch the analysis
needed to characterize the conformations, and elucidate the phase
behaviour mainly through average cluster properties.
Section~\ref{results} presents our numerical results, while
section~\ref{conclusions} summarizes our conclusions.

\section{Model and simulation methods}
\label{model}

We restrict ourselves to the most symmetric case of two-component
bottle-brush polymers with flexible backbones, i.e., side chains
composed of different monomers denoted as A and B have the
same length ($N_A=N_B=N$), with  $N$ varying within the range $N = 5 - 40$,
as is done in previous works~\cite{Theodorakis2011c,
Theodorakis2011d, Theodorakis2010b}. Although such range of chain lengths
may be short to compare with theoretical predictions, we emphasize
that this range of $N$ values corresponds nicely to the lengths used
in experiment~\cite{Zhang2005, Sheiko2008, Rathgeber2005,
Zhang2006, Hsu2010}.However, if one interprets a bead of the
flexible backbones in the bead-spring model as a Kuhn segment,
the simulated side groups may be longer in some of the cases
studied here than in real systems with the same number of monomers
per side group. Still, as an interesting theoretical, albeit
academic problem, it would be nice to extend our research efforts
to side-chain lengths of several hundreds of effective monomers,
but this is prohibitively difficult in the poor solvent regime,
where relaxation times of the chains are extremely long. The side
chains are grafted regularly and alternately onto a flexible
backbone with grafting density $\sigma$. There is one important
distinction comparing to previous work on bottle-brushes with
rigid backbones and those with flexible
backbones~\cite{Theodorakis2011, Theodorakis2012, Theodorakis2009,
Theodorakis2010, Theodorakis2011c, Theodorakis2011d,
Theodorakis2010b, Theodorakis2013}. For bottle-brush polymers with
flexible backbones the grafting density $\sigma$ is quantized: we
denote here by $\sigma=1.0$ the case that every backbone monomer
carries a side chain, $\sigma=0.5$ means that every second
backbone monomers has a side chain, etc. Backbone chain lengths
were chosen as $N_b=50$ and $N_b=100$. Backbone lengths longer
than $N_b=100$ are prohibitively difficult to study. According to
our previous experience, even bottle brushes in the $\Theta$
and good solvent regime~\cite{Theodorakis2011} with longer backbone
backbones are far from trivial to simulate.

We place a bottle-brush polymer in a simulation box with periodic
boundary conditions and dimensions such that no interaction of the
bottle-brush polymer with its periodic images could occur. We
describe both the backbone chain and the side chains by a
bead-spring model~\cite{Grest1986,Murat1989,Binder1995}, where all
beads interact with a truncated and shifted Lennard-Jones (LJ)
potential $U_{\rm LJ}(r)$ and nearest neighbors, bonded together
along a chain, also experience the finitely extensible nonlinear
elastic (FENE) potential $U_{\rm FENE}(r)$, $r$ being the distance
between the beads. Thus,
\begin{equation}
\label{Eq1} U_{\rm LJ} (r)=4 \epsilon_{\rm LJ}\left[ \left(
\frac{\sigma _{\rm LJ}}{r}\right)^{12}-\left(\frac{\sigma _{\rm
LJ}}{r}\right)^6 \right] +{\rm C}, \qquad r \leq r_{c},
\end{equation}
while $U_{\rm LJ}(r>r_c)=0$, and $r_c=2.5 \sigma_{\rm LJ}$. The
constant $\rm C$ is defined such that $U_{\rm LJ}(r=r_c)=0$ is
continuous at the cut-off. Henceforth, units are chosen such that
$\epsilon_{\rm LJ}=1$, $\sigma_{\rm LJ}=1$, the Boltzmann constant
$k_B=1$, and also the mass $m_{\rm LJ}$ of all beads is chosen to
be unity. When we consider two types (A, B) of side chains we
still use $\sigma^{AA}_{LJ}=\sigma^{BB}_{LJ}=\sigma^{BB}_{LJ}=1$
and $\epsilon^{AA}_{LJ}=\epsilon^{BB}_{LJ}=1$, but
$\epsilon^{AB}_{LJ}=0.5$ to create an un-mixing tendency between A
and B monomers. We know that in the case of a binary system with
monomers at density $\rho=1$ (i.e., a LJ mixture which is a
standard system for the study of phase separation), macroscopic
phase separation occurs below a critical temperature $T_c \approx
1.5$~\cite{Das2003}. For our bottle-brush polymers the average
densities are much smaller, especially at distances far from the
backbone, but since the critical temperature scales proportional
to the chain length, we are able to detect micro-phase separation
with our model. Moreover, due to the fact that the backbone is
fully flexible, bottle brushes adopt globular conformations under
poor solvent conditions favoring further macro-phase separation
compared to bottle brushes with rigid
backbones~\cite{Theodorakis2009, Theodorakis2013}.

The potential of equation~(\ref{Eq1}) acts between any pair of
beads, irrespective of whether they are bonded or not. For bonded
beads also the potential $U_{\rm FENE}(r)$ acts, where
\begin{equation}\label{Eq2}
 U_{\rm FENE}(r)=-\frac{1}{2} k r_{0}^{2}\ln\left[1-\left(\frac{r}{r_{0}}\right)^{2}\right]
\qquad 0<r\leq r_{0},
\end{equation}
$r_{0}=1.5$, $k=30$, and $U_{\rm FENE}(r)=\infty$ outside the
range written in equation~(\ref{Eq2}). Hence $r_0$ is the maximal
distance that bonded beads can take. Note that we did not include
any explicit solvent particles; solvent-mediated interactions and
solvent quality are only indirectly simulated by varying the
temperature of the system as is usually done~\cite{Binder1995}.

In our model the interactions of the side-chain monomers (A- and
B-type) with the backbone beads (denoted with the letter C here)
are the same [$\sigma^{\{A,B\},C}_{LJ}=1.0$,
$\epsilon^{\{A,B\},C}_{LJ}=1.0$]. Hence, the influence of the
backbone beads the phase behaviour of bottle-brushes is neutral,
given also that all the other parameters are chosen symmetrically.
This implies that, the polymer forming the backbone and the
side-chain polymers on coarse-grained scale are no longer
distinct. There is also no difference between the bond linking the
first monomer of a side chain to a monomer of the backbone and
bonds between any other pairs of bonded monomers. Of course, our
attempt here does not address any effects due to a particular
chemistry relating to the synthesis of these bottle-brush
polymers, but, as usually done~\cite{Binder1995}, we address
universal features of the conformational properties of these
macromolecules.

For the model defined in equations~(\ref{Eq1}) and (\ref{Eq2}),
the $\Theta$ temperature is known rather roughly~\cite{Grest1993},
namely, $\Theta \approx 3.0$. Being interested in $T \leq \Theta$,
we simulated the system for temperatures in the range $1.5 \leq T
\leq 3.0$. Note however that, equilibration of collapsed chains in
this temperature regime is considerably difficult, as will be
discussed below. In our simulations, the temperature was
controlled by the Langevin thermostat, following previous
work~\cite{Binder1995,Murat1991,Theodorakis2011b,Theodorakis2011e,
Theodorakis2011f, Theodorakis2012b,Murat1991}. The equation of
motion for the coordinates $\{r_i(t)\}$ of the beads
\begin{equation}\label{Eq3}
 m\frac{d^{2}\textbf{r}_{i}}{dt^{2}}=-\nabla U_{i}-
\gamma \frac{d\textbf{r}_{i}}{dt}+\Gamma_{i} (t)
\end{equation}
is numerically integrated using the GROMACS
package~\cite{Berendsen1995,Lindahl2001}. In equation~(\ref{Eq3}),
$t$ denotes the time, $U_{i}$ is the total potential acting on the
$i$-th bead, $\gamma$ is the friction coefficient, and
${\Gamma}_i(t)$ is the random force. As it is well-known, $\gamma$
and $\Gamma$ are related via the usual fluctuation-dissipation
relation
\begin{equation}\label{Eq4}
<\Gamma_{i}(t)\cdot
\Gamma_{j}(t^{'})>=6k_{B}T\gamma\delta_{ij}\delta(t-t^{'}).
\end{equation}
Following references~\cite{Binder1995,Grest1993,Murat1991}, the
friction coefficient was chosen as $\gamma=0.5$.
Equation~(\ref{Eq3}) was integrated with the leap-frog
algorithm~\cite{vanGunsteren1988} using an integration time step
of $\Delta t=0.006 \tau$, where the molecular-dynamics time unit
is $\tau=(m_{\rm LJ}\sigma_{\rm LJ}^{2}/\epsilon_{\rm
LJ})^{1/2}=1$.

Firstly, the system was equilibrated at a temperature $T=3.0$
using simulations extending over a time range of $30\times 10^6
\tau$. To gather proper statistics, we used $500$ independent
configurations at this high temperature, as initial configurations
for slow cooling runs, where the temperature was lowered in steps
of $0.1$, and the system was simulated, at each temperature, for a
time of $5 \times 10^6 \tau$. The final configuration of each
(higher) temperature was used as starting configuration for the
next (lower) temperature. In this way, one can safely generate, at
low temperatures and intermediate values of grafting densities,
statistically independent configurations. Let us note here that,
standard molecular-dynamics simulations would typically not sample
phase space adequately at this low-temperature regime. However,
this problem was overcome here by carrying out this large number
of independent slow cooling runs (typically around $500$ runs).
The statistical accuracy of our results was also checked by
varying on the same footing the cooling speed and length of runs.
For temperatures above $T=2.0$ correlations are rather
insignificant, but for temperatures below $T=2.0$ this slow
cooling methodology is indispensable in order to get reliable
results in our chosen range of parameters. We note that
in this simulation protocol higher values of $\gamma$ may be
used in order to reduce the relaxation time of the chains as
we are only interested for the equilibrium final states of
each one of these $500$ final configurations, but we decided
to use the same value of $\gamma$ that was introduced in the
standard bead-spring model~\cite{Grest1986}.

\begin{figure*}
\begin{center}
\rotatebox{0}{\resizebox{!}{0.60\columnwidth}{%
  \includegraphics{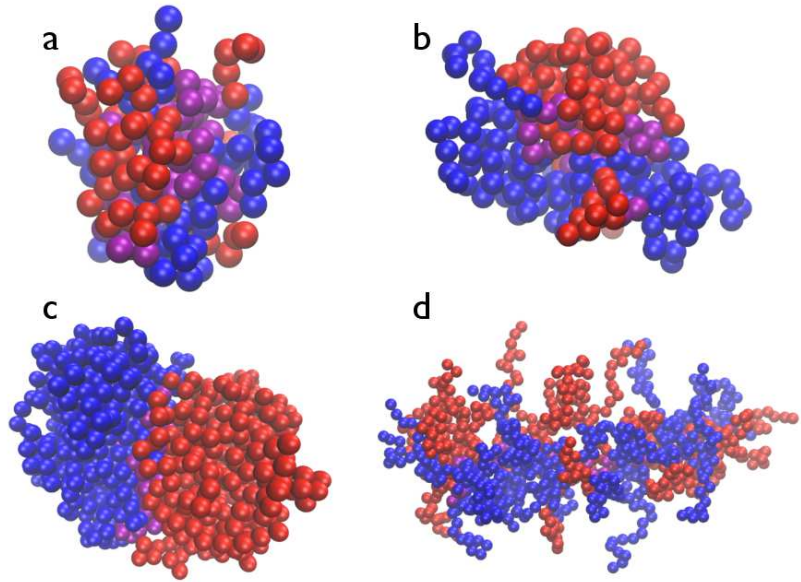}
}}
\end{center}
\caption{\label{fig1} (Colour online) Characteristic snapshots of
two component bottle-brush polymers with flexible backbones. Side
chain chains of type A/B are coloured as red/blue. The backbone
beads are shown in purple color. Parameters for these systems are
$\sigma=0.5$, $N_b=50$, $T=1.5$, and $N=5$ (a), $N=10$ (b) , and
$N=40$ (c). For case (d), $\sigma=1.0$, $N_b=50$, $N=20$, and
$T=3.0$. In panel (a), chains A and B can not fully phase
separate. In panel (b), macro-phase separation of A- and B-type of
monomers occurs with a certain probability $P(N_{cl}/M)$, which is
below unity. Bottle-brush polymers of case (c) are always phase
separated; therefore $P(N_{cl}/M)=1$. Panel (d) shows a snapshot
at $\Theta$ temperature in the brush regime. Here, side chains of
similar type are weakly correlated along the backbone resulting in
unstable cluster domains of A- or B-type of chains that
continuously fluctuate with regarding the number of chains that
they contain.}
\end{figure*}

In figure~\ref{fig1} we show typical configurations that provide a
visual overview of the phase behaviour of two-component
bottle-brush polymers with flexible backbones under poor solvent
conditions. At rather low temperatures far from the
$\Theta$-temperature (i.e., $T=1.5$) the polymer collapses into a
globular structure, where phase separation between side chains
occurs only partly, figures~\ref{fig1}(a) and (b), or fully,
figure~\ref{fig1}(c). The degree of phase separation depends
strongly on the side-chain length, as well as on the grafting
density. For very short side chains ($N=5$) phase separation never
takes place at a grafting density $\sigma=0.5$, whereas for $N=40$
macro-phase separation always occurs below a certain temperature
depending on the side-chain length $N$, and the grafting density.
As we will see below, the effect of the backbone length with
respect to the phase behaviour of bottle-brush polymer can be
considered as negligible.

When phase separation takes place, there is no reason a priori for
the system to decide which side chains will belong to a specific
cluster of the same type of monomers, also due to the symmetry of
our model. We actually observe fluctuations where a side chain of
type A or B that was part of a cluster of similar monomers escapes
from the cluster, to become part of another neighboring cluster
with chains of the same type of monomers. Such changes in the
number of chains per cluster occur rarely within the computational
time limits for partly phase separated systems like the one illustrated in
figure~\ref{fig1}(b). In fact, these fluctuations are the slowest
relaxation process in the system that lead to complete
thermodynamic equilibrium. As discussed above, in order to achieve
this equilibrium and perform a reliable statistical analysis we have
used the slow cooling runs. In this way, equilibrium cluster
distributions were obtained from our simulations. Without the
cooling runs procedure it would be impossible to run long
simulations that could be compared to the time scale of chain
exchanges between clusters.

In our analysis, we distinguished between clusters of type A and B
to investigate the phase separation in our model system. We
remind the reader that our symmetric choice of parameters for our systems
guarantee the same thermodynamic conditions for both types of chains. To
analyze the clusters, we have used the
Stillinger~\cite{Stillinger1963} neighborhood criterion for
monomers: if two monomers are less than a distance $r_n$ apart,
they belong to the same cluster, where
$r_n=1.5$.
Due to the symmetrically chosen set of
parameters for A- and B-type of monomers, the statistical analysis
for the relevant A and B clusters separately should give the same
outcome. This is confirmed by our results, validating our choice
of simulation procedure that allows us to obtain
statistically independent configurations at sufficiently low
temperatures.

\section{Results and discussion}
\label{results}

The phase behaviour of two-component bottle brushes can be
analyzed through the cluster analysis discussed in the previous
section. The quantity that characterizes the degree of phase
separation between monomers of different type is the probability
that a number of clusters per side chain occurs. Due to the
symmetry of our model this probability would be equal for the two
different type of chains, i.e.,
$P^{A}(N_{cl}/M)=P^{B}(N_{cl}/M)=P(N_{cl}/M)$. The first extreme
case would be that each grafted side chain forms a separate
cluster (The number of clusters $N_{cl}$ equals the number of
grafted chains A or B, i.e., $M^A=M^B=M$) with probability
$P(N_{cl}/M)=1$. In this case a side chain never forms a contact
with a neighboring side chain of the same type, and the formation
of clusters containing many side chains of the same type does not
take place. In this case, there is no phase separation between A
and B chains. The other extreme case would be that all chains are
always forming one cluster probability $P(1/M)=1$, which then
corresponds to a fully phase separated system, i.e., we have one
cluster of type-A chains and another cluster with type-B chains
with a well defined interface between them. The backbone beads are
part of this interface in a fully phase separated configuration.

\begin{figure*}
\begin{center}
\subfloat[][]{
\rotatebox{270}{\resizebox{!}{0.50\columnwidth}{%
  \includegraphics{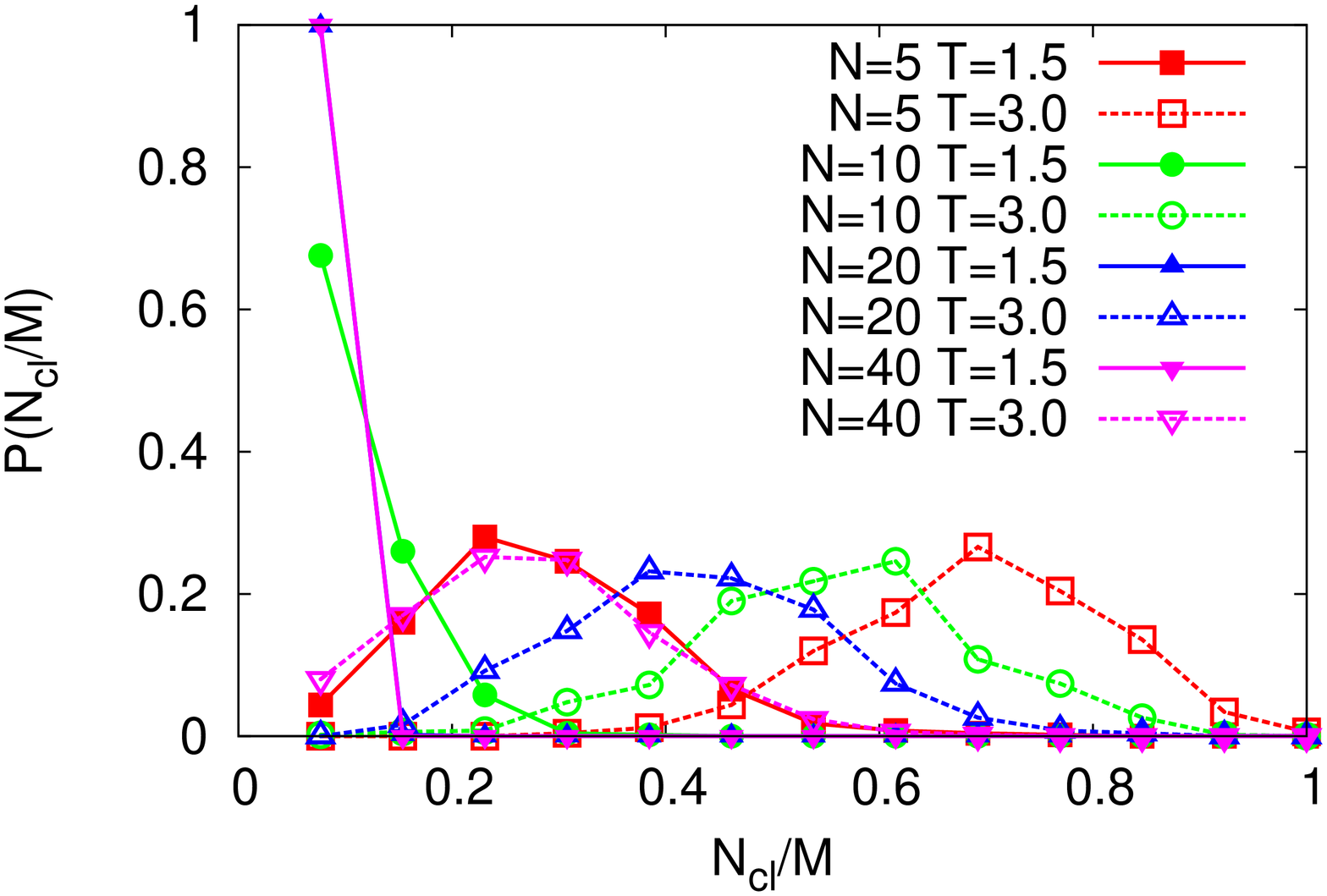}
}}} \subfloat[][]{
\rotatebox{270}{\resizebox{!}{0.50\columnwidth}{%
  \includegraphics{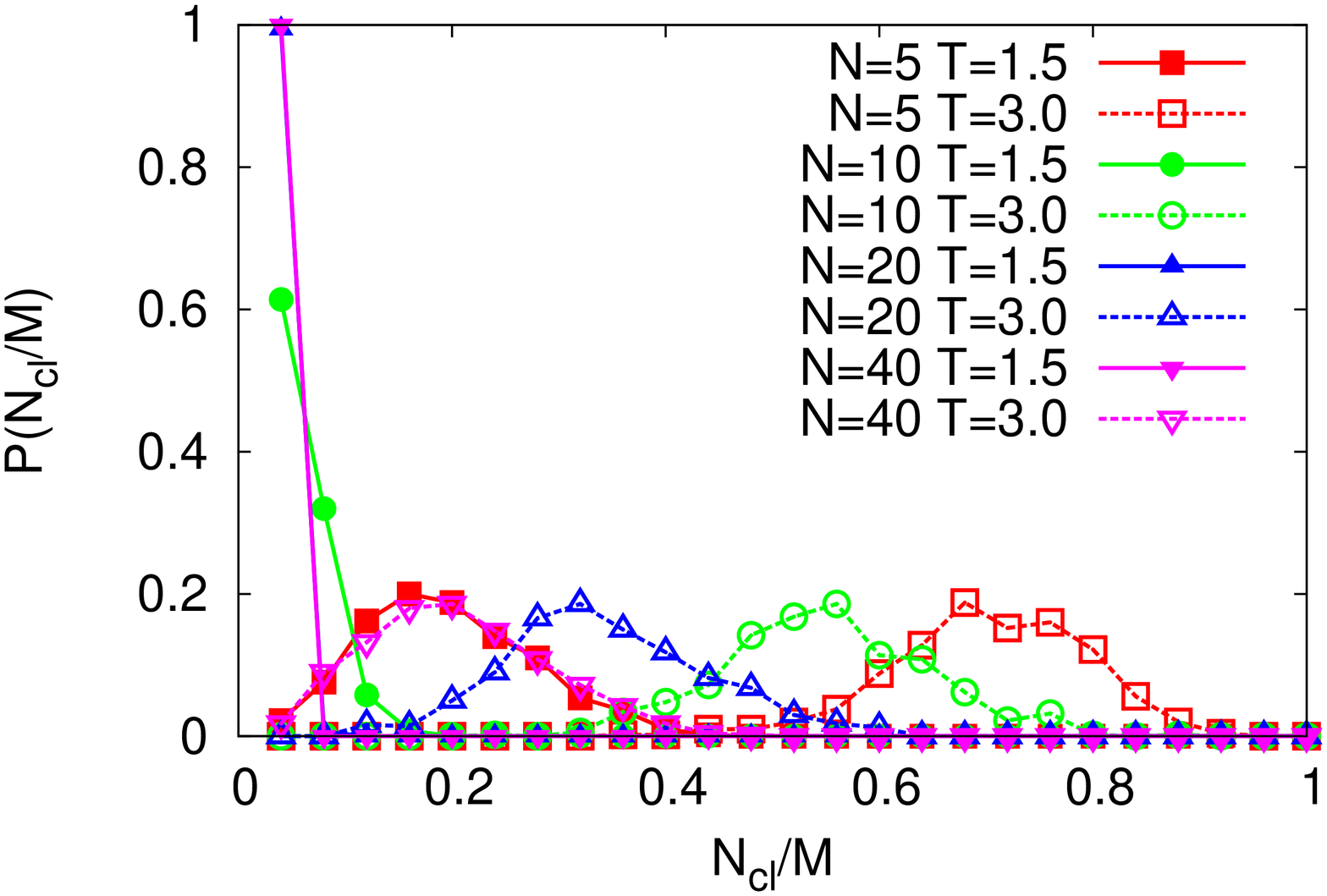}
}}}\\
\subfloat[][]{
\rotatebox{270}{\resizebox{!}{0.50\columnwidth}{%
  \includegraphics{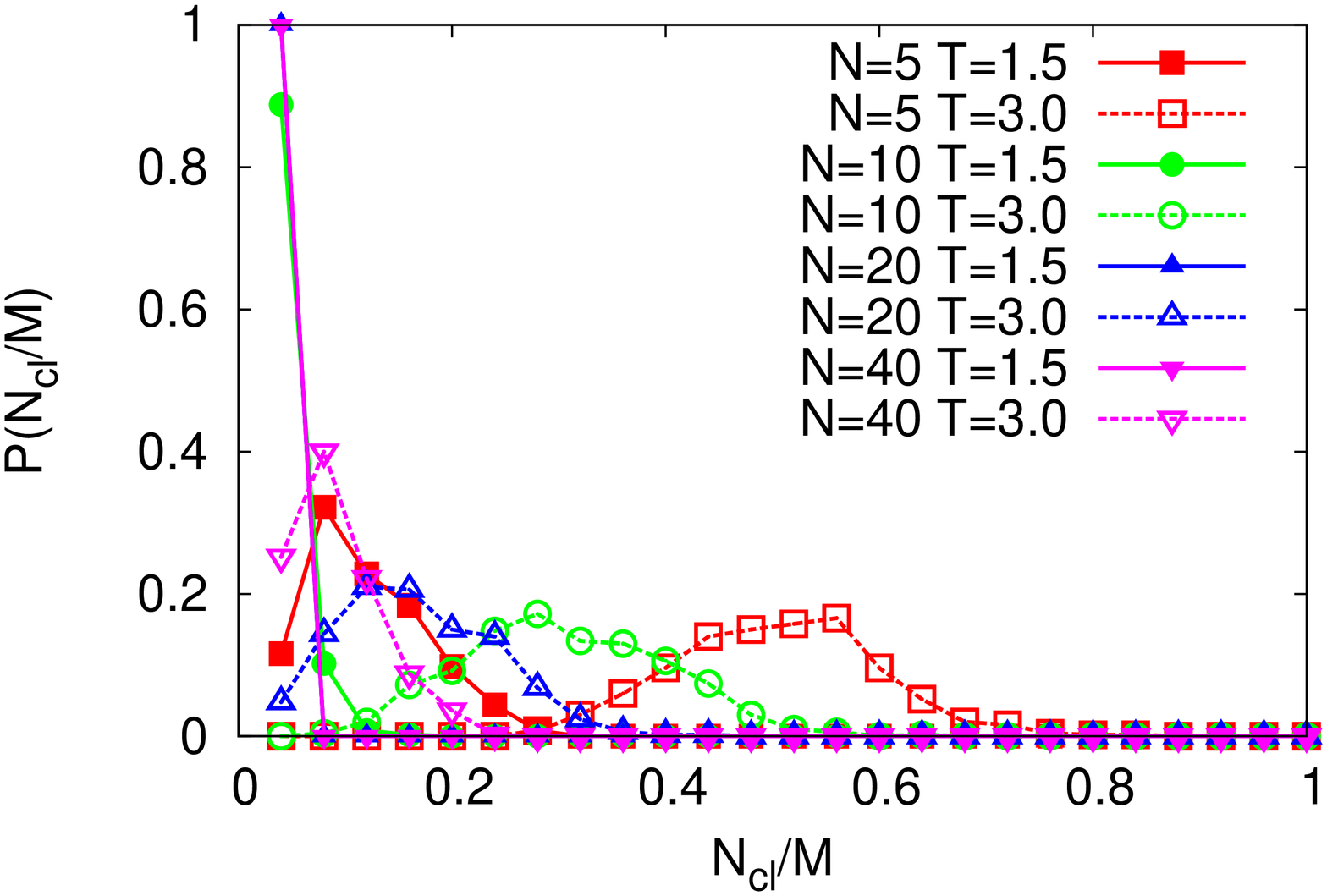}
}}} \subfloat[][]{
\rotatebox{270}{\resizebox{!}{0.50\columnwidth}{%
  \includegraphics{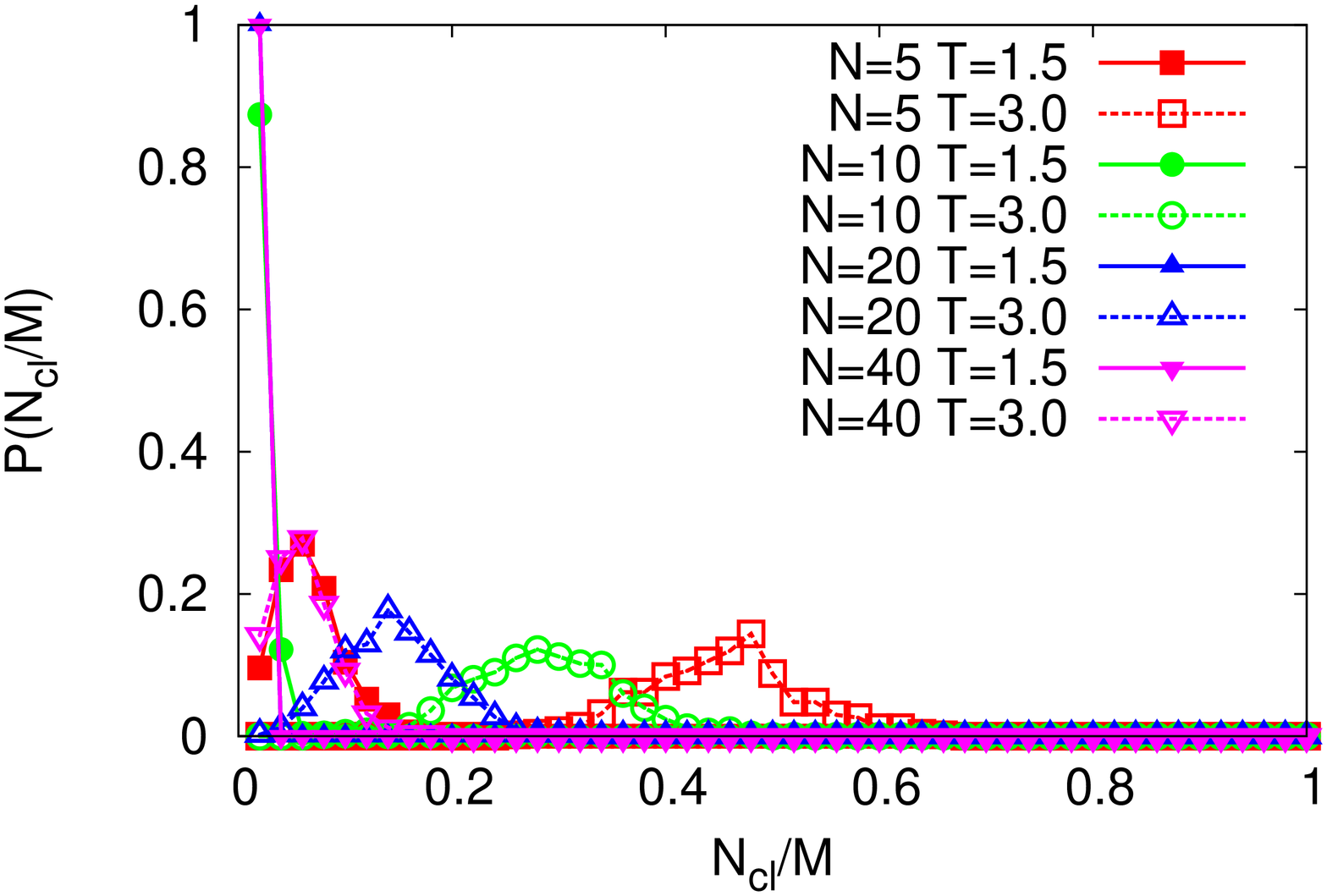}
}}}
\end{center}
\caption{\label{fig2} Probability $P(N_{cl}/M)$ that a number of
clusters $N_{cl}$ with type A or B side chains occurs. Due to the
symmetry of our model the analysis of A-type clusters gives the
same results as the analysis of clusters containing B-type chains.
Values in the horizontal axis are normalized to the total number
of grafted side chains (of type A or B) $M$ along the backbone. In
case (a) $\sigma=0.5$ and $N_b=50$, and the values of $N$ and $T$
are indicated on the graph. Similarly, in panel (b) we have
$\sigma=0.5$ and $N_b=100$. The results of panels (a) and (b)
suggest that the influence of the backbone length is negligible,
while the increase of the side chain length strongly favors the
phase separation of chains A from chains B. In panel (c)
$\sigma=1.00$ and $N_b=50$, whereas in (d) $N_b=100$. The increase
in the grafting density also favors macro-phase separation between
A- and B-type of side chains.}
\end{figure*}

In figure~\ref{fig2} we show the dependence of the probability
$P(N_{cl}/M)$ on the side-chain length $N$ for bottle-brushes of
various grafting densities and backbone lengths. In the graphs of
figure~\ref{fig2} we only show for clarity the lowest and the
highest temperatures that we considered in our study, namely $T=1.5$ and $T=3.0$. From
the results of figure~\ref{fig2} we see that the influence of the
backbone length $N_b$ is negligible, while the grafting density
$\sigma$ has a rather small effect in the range considered here.
The dependence on the backbone length $N_b$ is very weak;
practically, one may consider the influence due to the increase of
the backbone length negligible, whereas the increase of the
grafting density from $\sigma=0.5$ to $\sigma=1.0$ favors noticeably
formation of clusters. Moreover, the distribution in the size of
clusters becomes sharper as the grafting density increases showing
that fluctuations in the number of clusters that build the
cluster aggregations are becoming suppressed.
In figure~\ref{fig2}, we can also see that complete phase separation
between A and B chains for bottle-brushes with small side chains ($N=5$) is rarely
possible at high grafting densities. For $N=10$, phase
separation between A- and B-type of monomers occurs with
probability around $60\%$ for $\sigma=0.5$ and about $80\%$ for $\sigma=1.0$
for the lowest temperature ($T=1.5$). For side-chain lengths
longer than $N=10$, monomers A and B are always phase
separated. At temperatures close to the $\Theta$ ($T=3.0$) the
formation of clusters with A chains and clusters with B chains is
still possible even for short chain lengths.
Finally, the increase of the grafting density clearly
favors the formation of clusters at every temperature between
$T=1.5$ and $T=3.0$ as we will also discuss more in detail below.

\begin{figure*}
\begin{center}
\subfloat[][]{
\rotatebox{270}{\resizebox{!}{0.50\columnwidth}{%
  \includegraphics{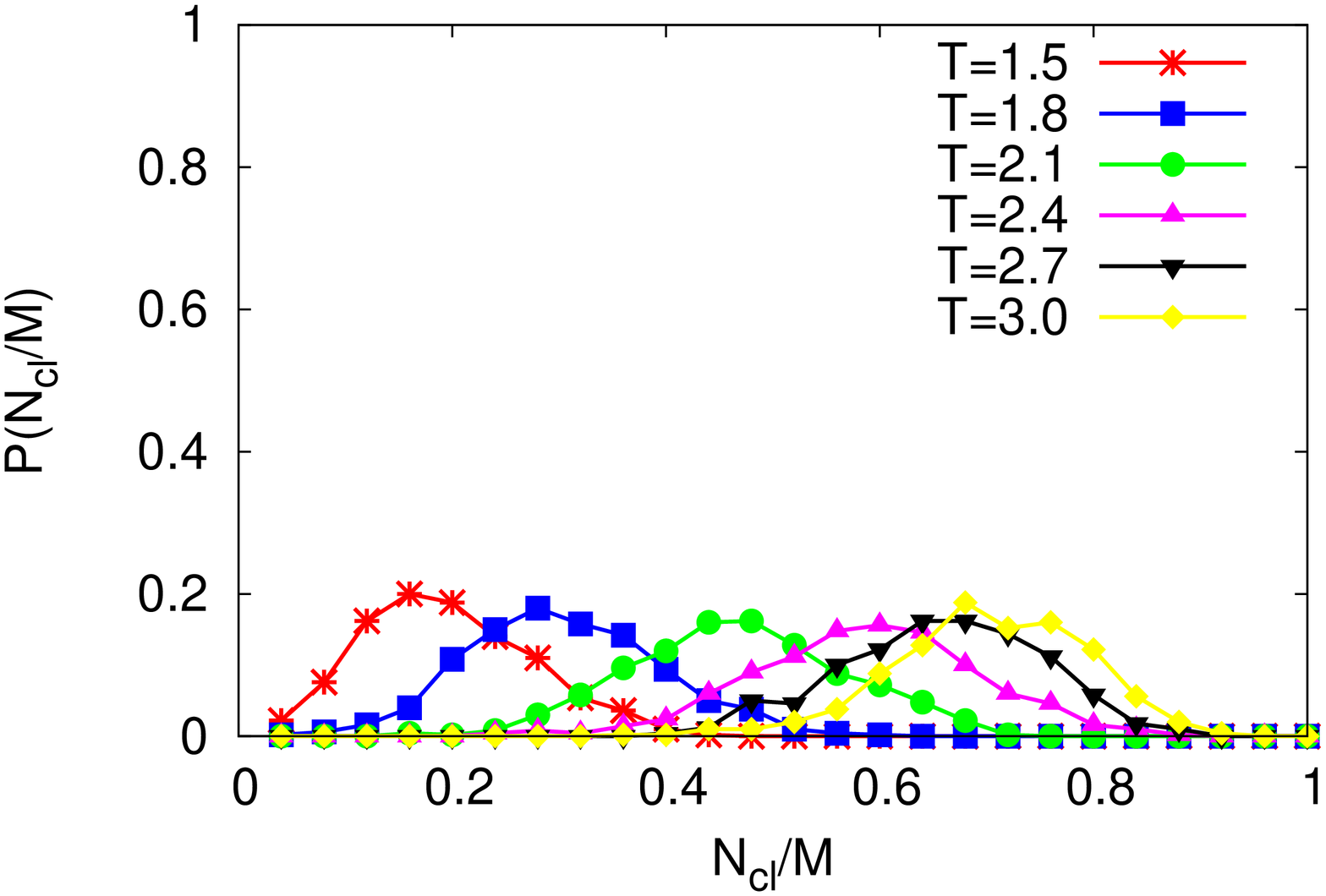}
}}} \subfloat[][]{
\rotatebox{270}{\resizebox{!}{0.50\columnwidth}{%
  \includegraphics{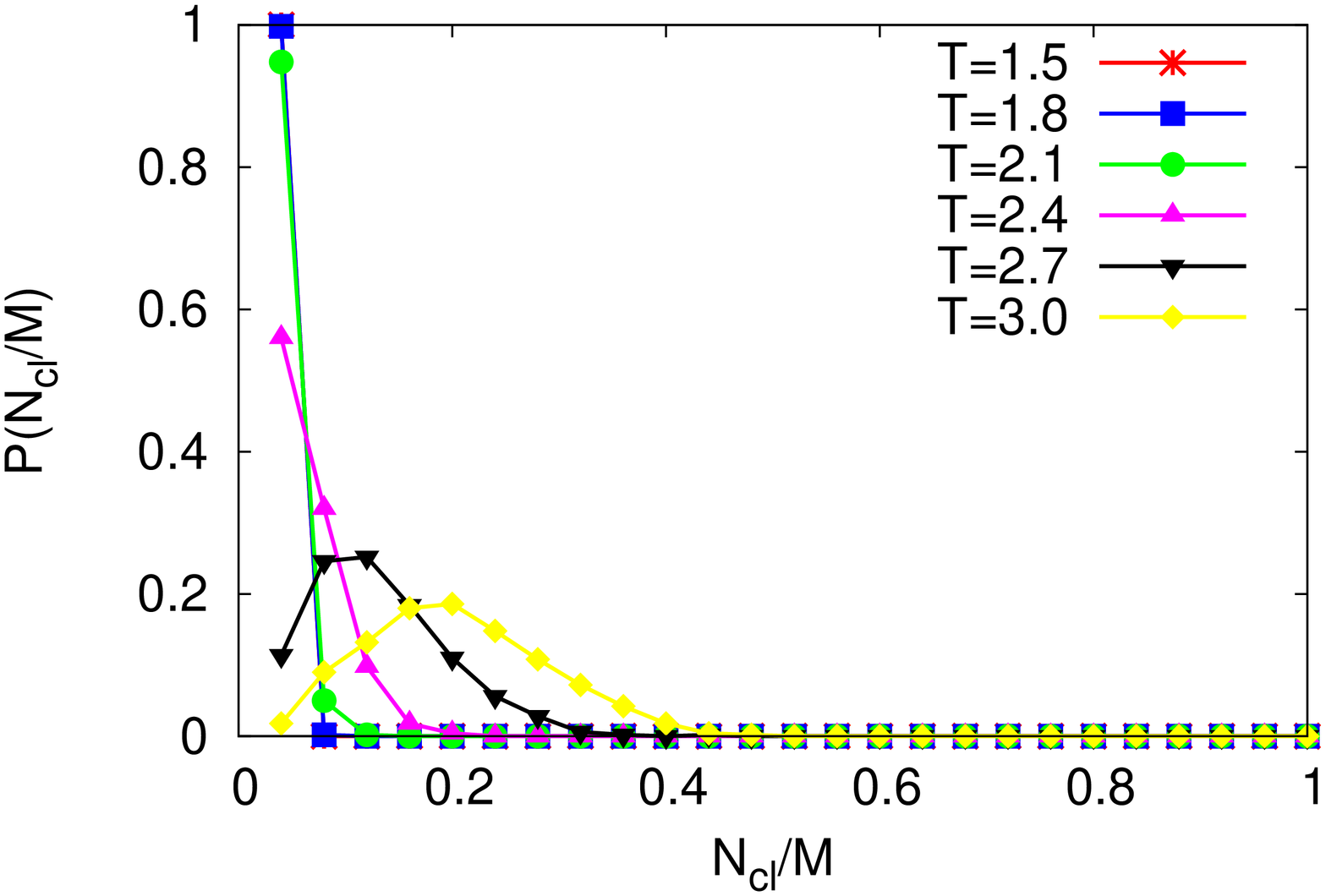}
}}}\\
\subfloat[][]{
\rotatebox{270}{\resizebox{!}{0.50\columnwidth}{%
  \includegraphics{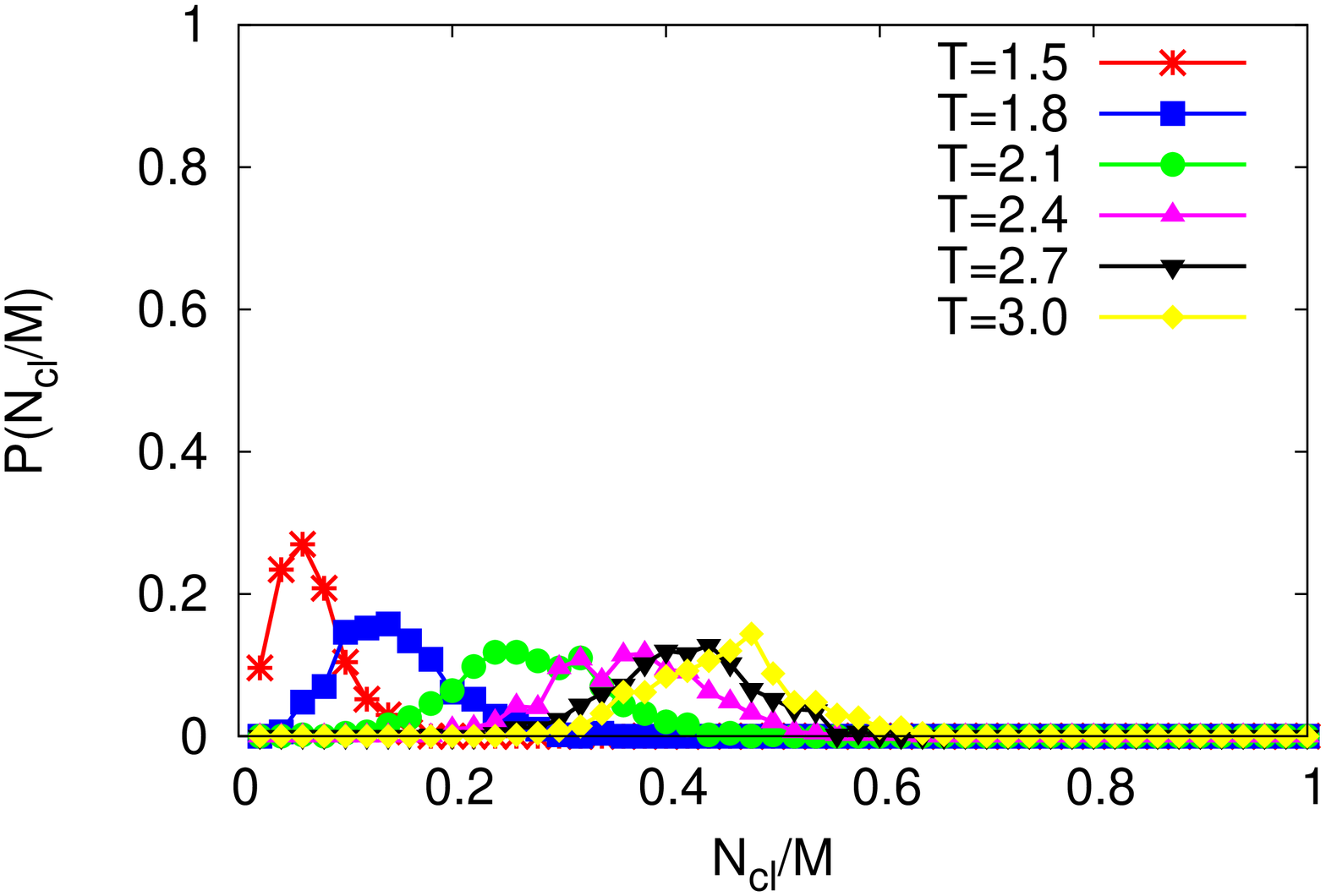}
}}} \subfloat[][]{
\rotatebox{270}{\resizebox{!}{0.50\columnwidth}{%
  \includegraphics{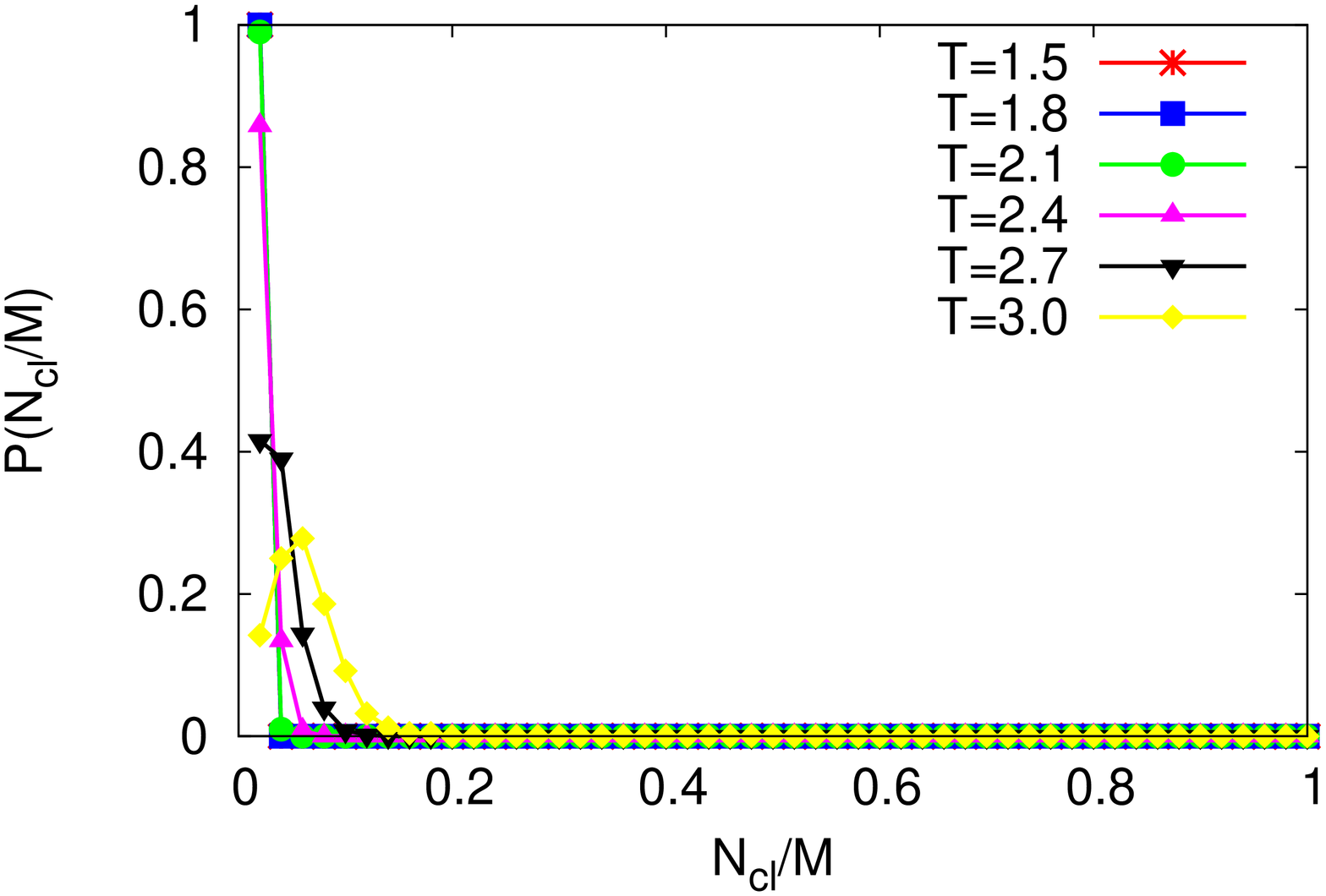}
}}}
\end{center}
\caption{\label{fig3} The dependence of the probability
$P(N_{cl}/M)$ on the temperature $T$ is shown for four
characteristic cases (a)-(d). We start with panel (a) where
$\sigma=0.5$, $N_b=100$, and $N=5$. For the range of temperatures
$T=1.5$ to $T=3.0$ clusters containing a few side chains are
formed. The chains that are part of a cluster can leave one
cluster and become part of another. The cluster sizes are
distributed symmetrically around an average value, as in
figure~\ref{fig2}. In this case, the grafting density and the
side-chain length are not high enough to lead to a fully
phase-separated system, where side-chains of type A are part of
single cluster and those of type B of another. In panel (b),
$\sigma=0.5$, $N_b=100$, and $N=40$. Here the side-chain length is
high enough to lead to a fully phase separated system for
temperatures $T<2.1$. A small variation in the number of clusters
is observed for the rest of the temperatures shown. Panel (c)
presents results for the case $\sigma=1.0$, $N_b=100$, and $N=5$.
Increase in the grafting density from $\sigma=0.5$ to $\sigma=1.0$
occasionally leads to a fully macro-phase separated system at
temperature $T=1.5$, although the side-chain length is still
short. Finally, panel (d) refers to the parameters $\sigma=1.0$,
$N_b=100$, and $N=40$. Here, chains of A- and B-type are for all
configurational samples fully separated at temperatures $T<2.4$.}
\end{figure*}

In figure~\ref{fig3} we illustrate the dependence of the cluster
formations on the temperature for four characteristic cases. In
figure~\ref{fig3}(a) no macro-phase separation takes place for the
whole range of temperatures between $T=1.5$ and $T=3.0$
($\sigma=0.5$ and $N_b=100$). As the temperature decreases the
formation of clusters is favored. Despite the short length of the
side chains, small clusters containing a few side chains are
possible even at $T=3.0$. When the side-chain length increases
[figure~\ref{fig3}(b)], i.e., $N=40$, macro-phase separation is
already initiated at temperature $T=2.7$. Moreover, at
temperatures below $T=2.1$ a cluster containing all side chains of
type A and another cluster containing all side chains of type B is
formed, with a well-defined interface between them that contains
the backbone monomers. When the grafting density increases
[figures~\ref{fig3}(c) and (d)], macro-phase separation is favored
at all temperatures. For $N=5$, figure~\ref{fig3}(c), macro-phase
separation occurs with a small probability for $\sigma=1.0$,
whereas never occurred for $\sigma=0.5$. The same is true for
$N=40$, comparing figures~\ref{fig3}(b) and (d). By examining all
the cases studied here, we found that at temperature $T=3.0$, no
macro-phase separation occurs for any of our cases; as expected
phase separation is initiated at temperatures below the
$\Theta$ temperature as a result of the favoring attractions
between beads of the same type.

Although the results of figures~\ref{fig2} and~\ref{fig3} clearly
provide an overview of the phase behaviour of bottle-brush
macromolecules with flexible backbones for the current range of
parameters, little information is provided on the interface
between phase separated systems. Therefore, we have also measured
the number of unfavorable contacts $n_{AB}$ for our systems which
is expressed by the following formula
\begin{equation}
\label{Eq5}n_{AB} = 4 \pi\int\limits_{0}^{r_{n}} {g}_{AB} (\Delta
r) (\Delta r)^{2}d(\Delta r),
\end{equation}
where $\Delta r$ is the absolute value of the distance between two
sites of monomers at positions $\vec{r}_{i},\vec{r}_{j}$, and
$g_{AB}$ the corresponding radial distribution function.
Equation~(\ref{Eq5}) means that a pair of monomers $(A, B)$ is
defined to have a pairwise contact if their distance is less than
$r_{n}$.

\begin{figure*}
\begin{center}
\subfloat[][]{
\rotatebox{270}{\resizebox{!}{0.50\columnwidth}{%
  \includegraphics{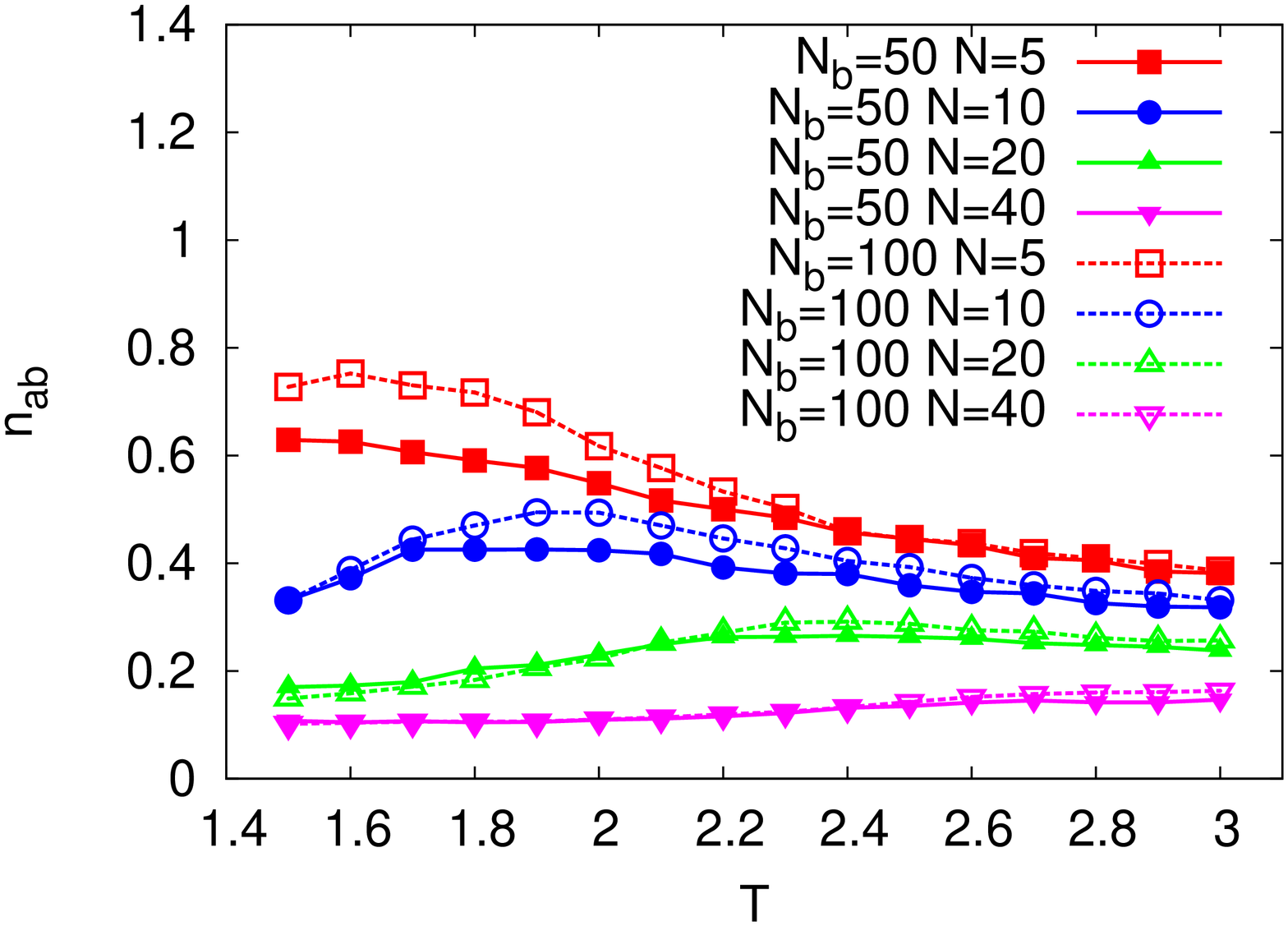}
}}} \subfloat[][]{
\rotatebox{270}{\resizebox{!}{0.50\columnwidth}{%
  \includegraphics{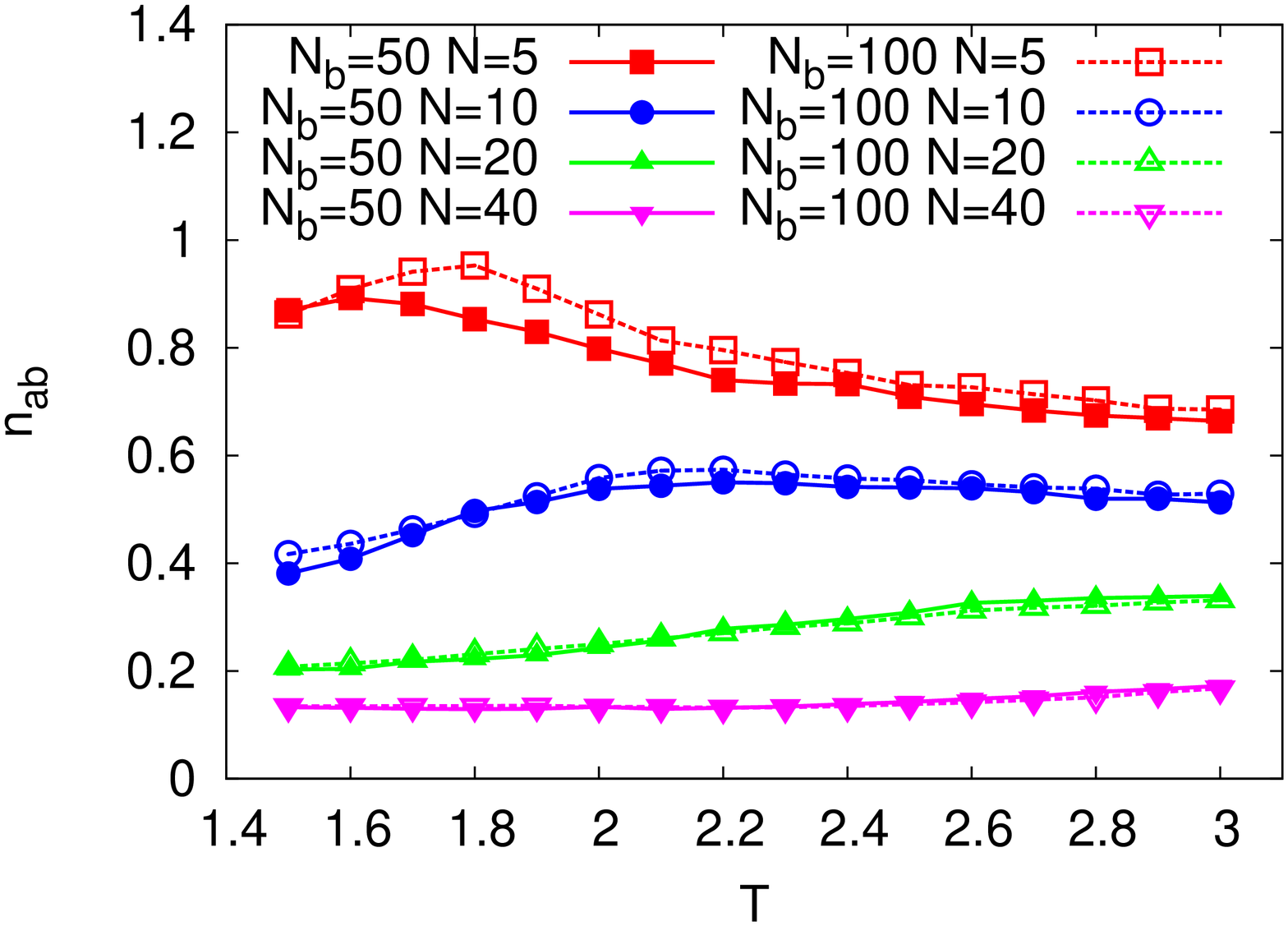}
}}}
\end{center}
\caption{\label{fig4} The number of unfavourable contacts, i.e.,
the number of contacts between A- and B-type monomers, as a
function of temperature for various bottle-brush polymers, as
indicated on the graphs. In panel (a) $\sigma=0.5$, while in panel
(b) $\sigma=1.0$. The different behaviour shown here is attributed
to the presence of the interface between A and B side chains in a
fully phase separated system [see, i.e., figure~\ref{fig1}(c)].
For more details, we refer to the discussion in the main body of
the text.}
\end{figure*}

Figure~\ref{fig4}(a) presents the average contacts per monomer
between monomers of A- and B-type for bottle brushes with grafting
density $\sigma=0.5$, whereas the results presented in
figure~\ref{fig4}(b) correspond to the case $\sigma=1.0$. The
results of figures~\ref{fig2} and~\ref{fig3} help us interpret
those of figure~\ref{fig4}, because they sketch the phase
behaviour of our systems. Overall, the increase of the grafting
density leads to a relevant increase in the number of contacts.
For both grafting densities $\sigma=0.5$ and $\sigma=1.0$ we can
distinguish three cases. In the first case, the bottle-brush
polymers with very short chains ($N=5$) and lower grafting density
($\sigma=0.5$) belong; here, macro-phase separation does not take
place and a well-defined interface between A- and B-type monomers
is not present. For this case a monotonic decrease in the number
of unfavorable contacts $n_{AB}$ occurs. At low temperatures,
where the bottle-brush has a globular structure, the increase of
the backbone length results in higher number of contacts between A
and B, which is a direct consequence of the increase in the number
of grafted side chains.

The second case corresponds to bottle-brush polymers for which at
lower temperatures only two clusters of A or B chains are always
fully separated with a well-defined interface, and at higher
temperature more clusters of A or B chains exist. At lower
temperatures an increase of $n_{AB}$ with the increase in
temperature is seen, as the attraction between beads of the same
type weakens and A- and B-type of monomers approach the interface.
At higher temperatures, the number of unfavorable contacts
$n_{AB}$ decreases with an increase in the temperature. Therefore,
the change in the slope as the temperature increases from lower to
higher temperatures signals the change in the phase behaviour of
the system and the disappearance of the well-defined A-B interface
between A and B chains.

The third case we could identify from the graphs of
figure~\ref{fig4} corresponds to a ``strong'' macro-phase
separation between A- and B-type of monomers for a large range of
temperatures. These are cases that are related to bottle-brush
polymers with high grafting density and, more importantly, long
side chains. Here, the number of unfavorable contacts $n_{AB}$
shows a monotonic increase with the increase of the temperature.
For relatively long side chains phase separation between
individual chains becomes strong and an interface between
individual chains is possible at temperatures below the $\Theta$
temperature. Therefore, we see a considerable change in behaviour
between bottle-brushes with short side chains and bottle-brushes
with long side chains. In the latter case, phase separation
between individual grafted chains becomes pronounced.

\begin{figure*}
\begin{center}
\subfloat[][]{
\rotatebox{270}{\resizebox{!}{0.50\columnwidth}{%
  \includegraphics{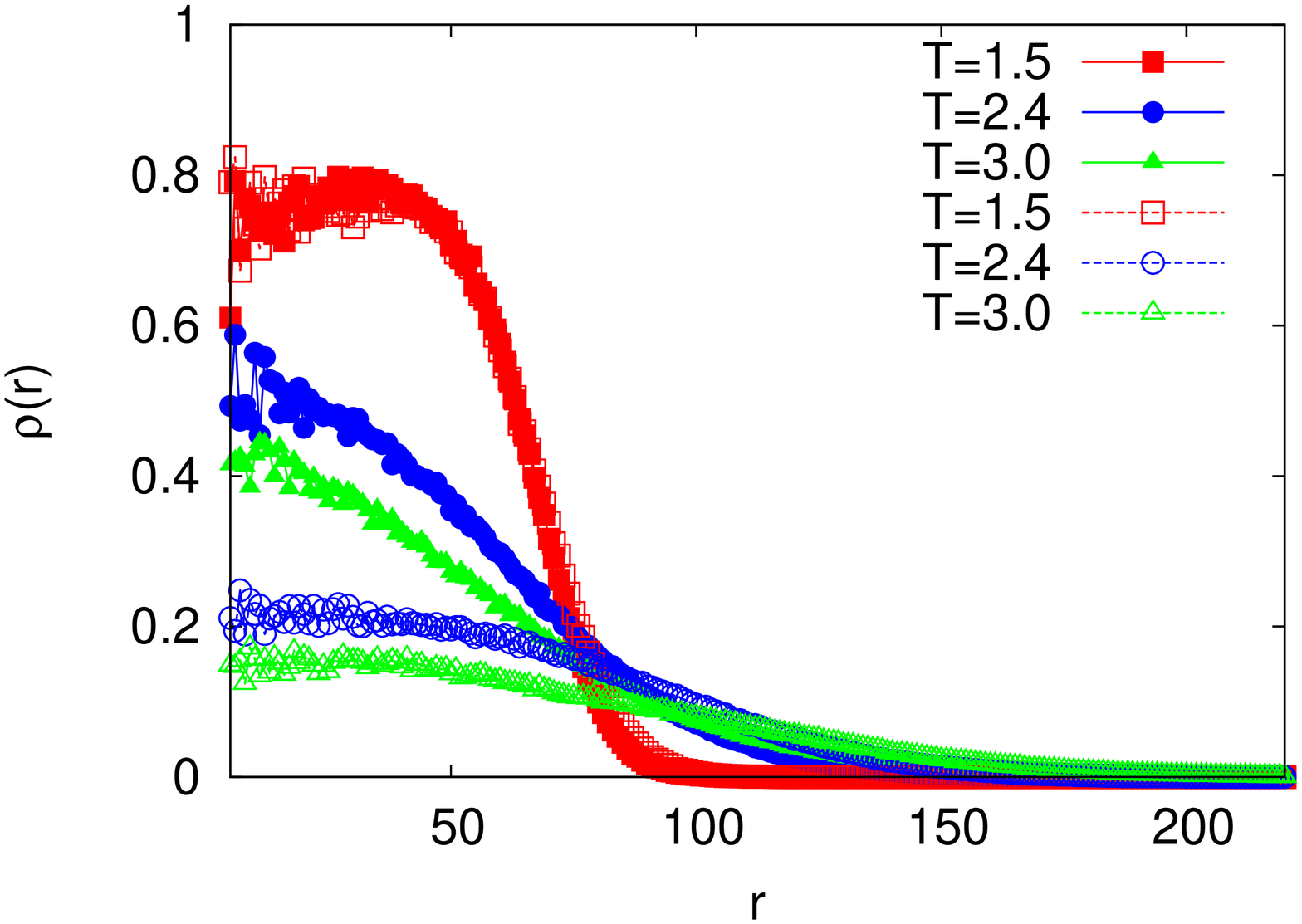}
}}}\\
\subfloat[][]{
\rotatebox{270}{\resizebox{!}{0.50\columnwidth}{%
  \includegraphics{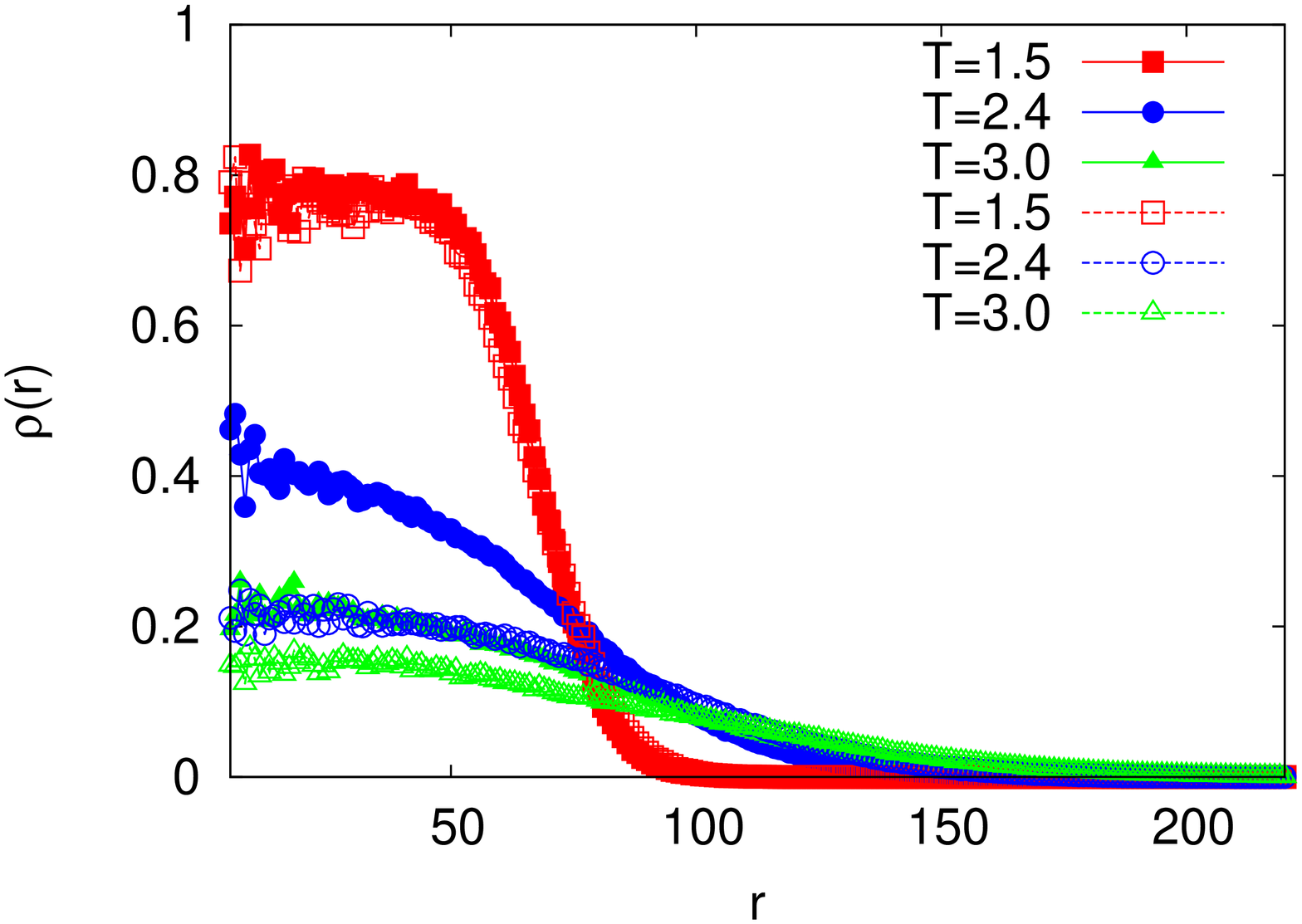}
}}}\\
\subfloat[][]{
\rotatebox{270}{\resizebox{!}{0.50\columnwidth}{%
  \includegraphics{fig5b.eps}
}}}
\end{center}
\caption{\label{fig5} Density profiles $\rho(r)=\rho(|\vec{r}|)$
plotted versus the radial distance $r=|\vec{r}|$ from the center
of mass of the bottle-brush. Various temperatures are shown as
indicated. Case (a) presents results for bottle-brush polymers
with $\sigma=1.0$, $N_b=50$, and $N=20$ (full symbols) and
$\sigma=1.0$, $N_b=100$, and $N=10$ (open symbols). In graph (b)
results refer to bottle-brush macromolecules with $\sigma=0.5$,
$N_b=100$, and $N=20$ (full symbols) and $\sigma=1.0$, $N_b=100$,
and $N=10$ (open symbols). Panel (c) refers to $\sigma=0.5$,
$N_b=100$, and $N=40$ (full symbols), and $\sigma=1.0$, $N_b=100$,
and $N=20$ (open symbols). Whereas in the case of multi-block
copolymers the interface of a macro-phase separated system can be
easily detected through averaged overall
profiles~\cite{Theodorakis2011b, Theodorakis2011f}, this is
clearly not the case for the bottle-brush macromolecules of this
study, since only subtle differences in the density profiles
appear.}
\end{figure*}

The above analysis of the unfavorable contacts provide some clue
for the presence of a well-defined interface between A and B
chains. However, we could gain additional insight in the overall
picture of the bottle brushes examined in this work by computing
the averaged overall density of monomers, which is mathematically
expressed by the following formula
\begin{equation}\label{Eq6}
\rho(|\vec{r}|)= \langle  \sum \limits_{i=1}^{nN} \delta(
\vec{r}-\vec{r}_{c}-\vec{r}_{i}) \rangle.
\end{equation}
In the above equation $\delta(\vec{x})$ is the Dirac delta
function, $\vec{r}_c$ is the position of the center of mass of all
monomers that belong to the chain, and $\vec{r}_i$ are the
positions of all monomers, irrespective of whether they belong to
the backbone or to the side chains, and also irrespective of
whether they are of A or B type. The angle brackets denote an
average over all conformations, as usual.

Figure~\ref{fig5} shows three characteristic examples of the
overall monomer density for different choices of parameters
$\sigma$, $N_B$, and $N$, and for three different temperatures only
chose for clarity.
Cases on each graph of figure~\ref{fig5} correspond to two
different bottle-brushes that contain the same total number of
monomers. For example, figure~\ref{fig5}(b) presents results for
one bottle brush with parameters $\sigma=0.5$, $N_b=100$, and
$N=20$, and another with parameters $\sigma=1.0$, $N_b=100$, and
$N=10$. Similarly are chosen the parameters for graphs (a) and (c)
of figure~\ref{fig5}. From the figures we can see that
the overall monomer distribution of the monomers for macro-phase
separated systems at temperature $T=1.5$ is similar for different
bottle brushes with the same total number of monomers. Therefore,
it is not possible to verify from this property the presence of a
well-defined interface between A- and B-type of monomers. At
higher temperatures ($T=2.4$ and $T=3.0$) more pronounced
differences are observed in the density profiles of the bottle-
brushes depending on the grafting density $\sigma$ and the
side-chain length $N$. Examination of the profiles of
figure~\ref{fig5} can also provide further information on the
transition of a globular structure at low temperatures to a
bottle-brush-like structure at higher temperatures, i.e.,
temperatures close to the $\Theta$ temperature. The apparent
similarities in the density profiles and the fact that the
interface A-B does not play such a significant role as for
other polymer macromolecules of complex architecture
(e.g., multi-block co-polymers~\cite{Theodorakis2011b,
Theodorakis2011e, Theodorakis2011f, Theodorakis2012b})
may suggest that the conformational properties of two-component
bottle-brushes (e.g. mean square gyration radius of side chains, etc.)
may not differ significantly from the
corresponding properties for single-component bottle-brushes
with flexible backbones~\cite{Theodorakis2013}.

\section{Conclusions}
\label{conclusions}

Prior work on bottle-brush polymers has documented the phase
behaviour of bottle brushes with rigid backbones under poor
solvent conditions. It was found that for intermediate and high
grafting densities, pearl-necklace and Janus-like structures are
formed. In the present contribution, we discussed the phase
behaviour of two-component bottle-brush polymers with flexible
backbones under poor solvent conditions, by means of extensive
molecular-dynamics simulations.

For our range of parameters that can be directly compared to
bottle-brushes with a rigid backbone of previous
studies~\cite{Theodorakis2011c}, we found that the formation of
pearl-necklace or Janus-like structures does not take place when
the backbone is fully flexible. Therefore, the aforementioned
structures are totally attributed to the rigidity of the backbone.
When the grafting density, or the side-chain length, increases,
phase separation is favoured, indeed. The backbone length seems to
play no role for the range of parameters examined here. Our
findings extend previous work on bottle-brush polymers with rigid
backbones~\cite{Theodorakis2011c} and provide an outline on the
phase behaviour of two-component bottle-brush polymers, since the
fully flexible and rigid backbone cases are the two limiting cases
for this type of complex systems.

Thus far, all simulation studies have been realized for rather
moderate chain lengths, which, however, serve our current needs
for comparison with experimental work. As a future challenge, it
would be favourable to have numerical data for larger systems,
although this task is currently hindered by the huge relaxation
times of such macromolecules at low temperatures. Finally, it is
worth noting that a more detailed study for intermediate
rigidities that may reveal interesting structures would be
desirable, serving as a bridge between the current and previous
works~\cite{Theodorakis2010,Theodorakis2011c}.

\end{document}